\documentclass[12pt]{article}
\pdfoutput=1
\usepackage{color}
\usepackage{fullpage}
\usepackage{amsmath}
\usepackage{amssymb}
\usepackage{amscd}
\usepackage{amsthm}
\usepackage{graphicx}
\usepackage{amssymb}
\usepackage{epstopdf}
\usepackage{cancel}
\usepackage{setspace}
\usepackage{slashed}
\usepackage{subfigure}
\usepackage[hang,small]{caption}
\def\p{\partial}

\newcommand{\be}{\begin{equation}}
\newcommand{\ee}{\end{equation}}

\newcommand{\mcal}[1]{\mathcal{#1}}

\theoremstyle{definition}

\numberwithin{equation}{section}
\begin{document}
\begin{titlepage}
\bigskip
\rightline{}

\bigskip\bigskip\bigskip\bigskip
\centerline {\Large \bf {Black Funnels}}
\bigskip\bigskip

\centerline{\large  Jorge E. Santos and Benson Way}
\bigskip\bigskip
\centerline{\em Department of Physics, UCSB, Santa Barbara, CA 93106}
\centerline{\em jss55@physics.ucsb.edu, benson@physics.ucsb.edu}
\bigskip\bigskip
\begin{abstract}
The Hartle-Hawking state of $\mathcal{N}=4$ SYM at strong coupling and large $N$ on a fixed black hole background has two proposed gravitational duals: a black funnel or a black droplet. We construct the black funnel solutions that are dual to the Hartle-Hawking state on a Schwarzschild black hole and on a class of three-dimensional asymptotically flat black hole backgrounds.  We compute their holographic stress tensor and argue for the stability of these solutions.  
\end{abstract}
\end{titlepage}


\onehalfspacing

\begin{section}{Introduction}
A pivotal moment in the study of quantum fields on black hole backgrounds was the realization that black holes will evaporate by emission of Hawking radiation \cite{Hawking:1974rv}.  The fundamental issues associated with this discovery have played a central role in the quest for a quantum theory of gravity.  However, much of this understanding comes from the study of free field theories.  When the fields are strongly interacting, not much is known beyond a formal definition of a Hartle-Hawking state from a Euclidean path integral. 

More recently, the AdS/CFT correspondence \cite{Maldacena98,Gubser:1998bc,Witten:1998qj} has provided a means of studying certain strongly coupled field theories.  States of a large $N$, strongly coupled field theory in a background $\mcal M$ are mapped to solutions of a theory of gravity in one higher dimension with conformal boundary $\mcal M$.  Typically, the field theory background is Minkowski space.  In order to understand strongly coupled Hawking radiation, the authors in \cite{Hubeny:2009ru,Hubeny:2009kz,Hubeny:2009rc} instead applied the correspondence to a black hole background.  

What then, is the gravity dual to a Hartle-Hawking state?  From the perspective of the field theory, the stress tensor must approach that of a thermal fluid far from the boundary black hole.  Then in the bulk, the solution should approach that of a planar black hole (also known as a black brane).  Furthermore, there must be a horizon near the conformal boundary that matches up with the boundary black hole horizon.  Therefore, the authors of \cite{Hubeny:2009ru} conjecture the existence of two families of solutions:  the two horizons either join up to form a connected horizon,  or they are disconnected.  In the former, the solution forms a \emph{black funnel}, while in the latter, the solution is a \emph{black droplet} suspended over a (deformed) black brane, see Fig.~\ref{figs:sketches}.  Because of the connected horizon, the black funnels are expected to be dual to a situation where the boundary black hole couples to the field theory plasma.  By contrast, the black droplet solutions describe a boundary black hole that does not easily exchange heat or other quantities with the plasma.  
\begin{figure}[t]
	\centering
	\subfigure[Black Droplet]
		{
			\includegraphics[width=0.45\textwidth]{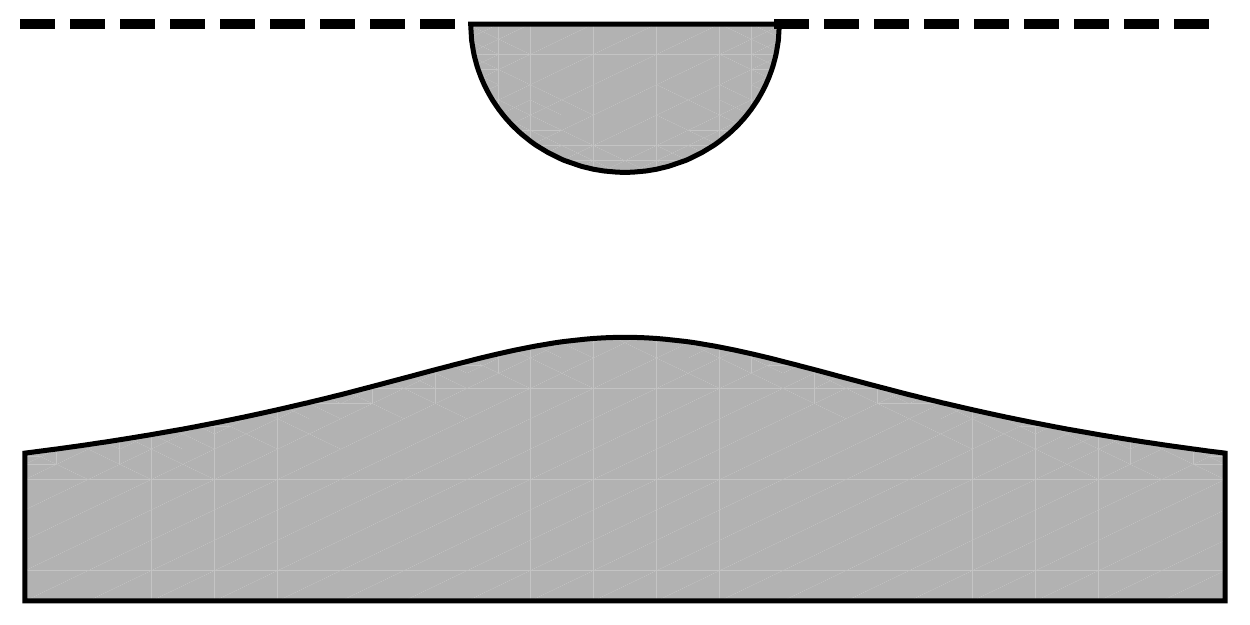}
			\label{fig:sketch1}
		}
	\subfigure[Black Funnel]
		{
			\includegraphics[width=0.45\textwidth]{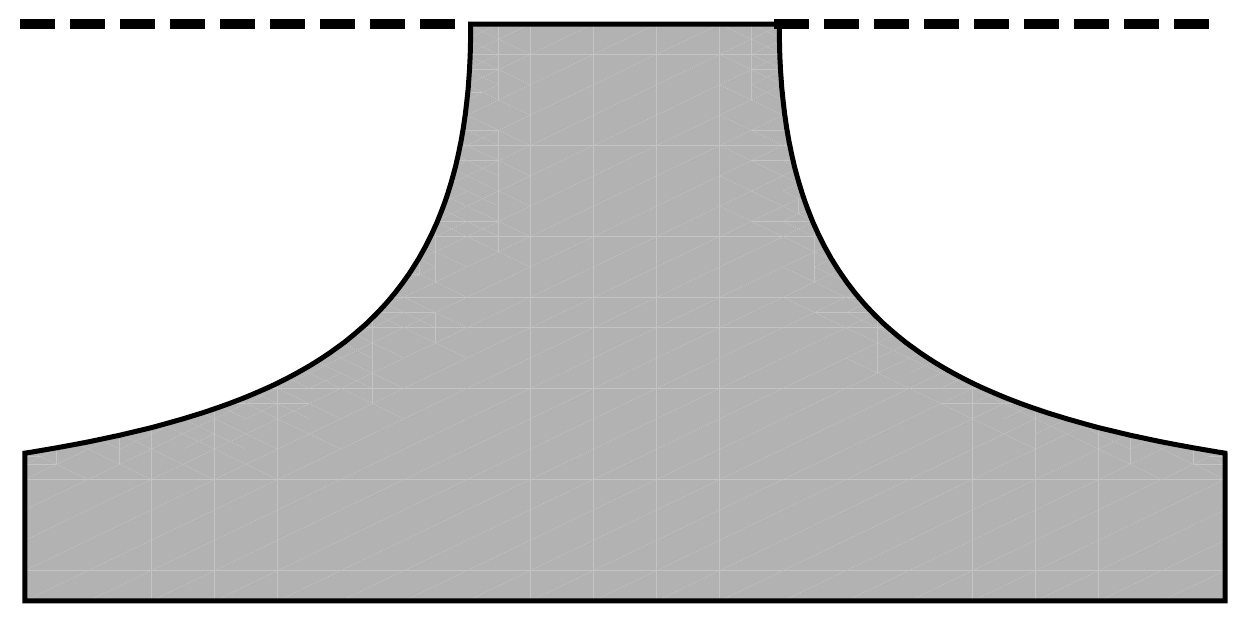}
			\label{fig:sketch2}
		}
	\caption{\label{figs:sketches}Sketches for black droplets and funnels.}
\end{figure}

A phase transition between these two phases was also conjectured in \cite{Hubeny:2009ru}.  If the product of the size of the boundary black hole and the temperature is small $RT\ll1$, then the black funnel may have a long narrow neck, and might become unstable to the Gregory-Laflamme instability \cite{Gregory:1993vy}.  It is therefore natural to suggest that black droplets dominate the ensemble for $RT\ll1$.  Conversely, if $RT\gg1$, the droplet horizon might extend deep into the bulk until it nearly touches the planar horizon.  It may become favourable for these horizons  to merge, and so the funnel phase might be preferred at $RT\gg1$.  

Various black funnel and black droplet solutions have been constructed from the AdS C-metric \cite{Hubeny:2009ru,Hubeny:2009kz,Caldarelli:2011wa}.  These solutions, however, do not have a boundary black hole that is asymptotically flat.  A black droplet dual to Schwarzschild was also found in \cite{Figueras:2011va}.  There, the solution does not have a planar black hole, and is therefore dual to an Unruh state\footnote{Note that a Boulware state would require a minimum energy configuration, which would be similar to the solution found in \cite{Figueras:2011va}, but with an extremal horizon.}.

In this paper, we numerically construct the first black funnel solutions with a boundary metric that is asymptotically flat.  We work in five dimensions where the bulk metric is dual to Schwarzschild, and also in four dimensions where we vary the size of the boundary black hole.  

We also compute the boundary stress tensor for these solutions.  However, in five bulk  dimensions there will be a conformal anomaly unless the boundary metric is Ricci flat.  This will generically introduce logarithms in the expression for the stress tensor, which pose a difficultly for our numerical method.  The four-dimensional case has no conformal anomaly, even for non-Ricci flat solutions.  It is chiefly for this reason that we do not vary the size of the $d=5$ black funnels. 

This paper is structured as follows.  In the next section, we describe the construction of these black funnel solutions and our numerical method.  In the following sections, we compute the stress energy tensor and study these solutions by embedding them isometrically in hyperbolic space.  We then finish with a few concluding remarks.  
\end{section}

\begin{section}{Constructing Black Funnels} 
\begin{subsection}{The DeTurck Method}
Our black funnels require only Einstein gravity and a cosmological constant.  The action in $d$ bulk spacetime dimensions is
\be
S=\int d^dx\;\sqrt{-g}\left(R-2\Lambda\right)\;,\qquad \Lambda=-\frac{(d-1)(d-2)}{2L^2}\;,
\ee
where $L$ is the AdS length scale.  Rather than solving the Einstein equation directly, we will solve the Einstein-DeTurck equation (also known as the harmonic Einstein equation):
\be\label{deturck}
R_{\mu\nu}=\frac{2\Lambda}{d-2}g_{\mu\nu}+\nabla_{(\mu}\xi_{\nu)}\;,
\ee
where $\xi^\mu=g^{\alpha\beta}\left(\Gamma^{\mu}_{\alpha\beta}-\bar\Gamma^{\mu}_{\alpha\beta}\right)$, and $\bar\Gamma^{\mu}_{\alpha\beta}$ is the Levi-Civita connection of some reference metric $\bar g$.  This reference metric is chosen to have the same asymptotic and horizon structure as our metric $g$.  For the metrics we are concerned with, $\eqref{deturck}$ is elliptic \cite{Headrick:2009pv}.  

In general, solutions to $\eqref{deturck}$ will not be solutions of the Einstein equation unless $\xi=0$.  We will sketch a proof that solutions with $\xi\neq0$, called Ricci solitons, do not exist, closely following \cite{Figueras:2011va}. We start by moving to the Euclidian section, setting the time direction of our Lorentzian manifold to be $-i\tau$. Note that because we are interested in static Lorentzian solutions, we can always do this. Our manifold, $(\mathcal{M},g)$ is now Riemannian and satisfies (\ref{deturck}). Before proceeding, let us mention that by taking the divergence of (\ref{deturck}) and using the Bianchi identities, one can show that $\chi \equiv \xi^\mu \xi_\mu\geq0$ satisfies the following differential equation:
\be
\nabla^2 \chi+ \xi^\mu \nabla_\mu \chi = -\frac{4 \Lambda}{d-2}\chi+2(\nabla^\mu\xi^\nu)(\nabla_\mu \xi_\nu)\geq 0,
\label{eq:chi}
\ee
where we noted that the right hand side of (\ref{eq:chi}) is positive semi-definite. In order for a solution of (\ref{eq:chi}) to exist, it is necessary that the associated equation
\be
\nabla^2 f+ \xi^\mu \nabla_\mu f\geq0
\label{eq:f}
\ee
has a solution for a function $f$ on a nontrivial background $(\mathcal{M},g,\xi)$ with $f\geq0$. Furthermore, the behavior of $f$ near the boundaries of $\mathcal{M}$ need to be the same as $\chi$. However, it is know \cite{Figueras:2011va} that Eq.~(\ref{eq:f}) admits a maximum principle, which for non-vanishing $f$ states the following: i) $f$ can only have a maximum on $\partial \mathcal{M}$, ii) at $\partial \mathcal{M}$, the outer normal derivatives of $f$, $\partial_n f$, must be positive semi-definite. In particular, if $f\geq0$ is zero on $\partial \mathcal{M}$, statement i) implies that $f$ must be everywhere zero. The authors of \cite{Figueras:2011va} have shown that $\chi$ vanishes for a locally asymptotically AdS boundary with reference metric such as the ones we are going to use, and also at horizons (extremal and non-extremal). Therefore, if we want to show that $\chi$ is \emph{everywhere} zero, we must ensure that $\chi$ is zero on all of our boundaries\footnote{We thank the referee for pointing this out to us in an earlier version of this manuscript.}.  We will demonstrate this later when we discuss the boundary conditions for our solutions.  

Once we can establish that no Ricci solitons exist, we can use $\xi$ to monitor our numerical error; we check that $\xi=0$ to machine precision. This method also has the advantage that a gauge choice need not be made a priori.  Solving for the equations will also choose the gauge $\xi=0$. 

\end{subsection}
\begin{subsection}{Integration Domain}
The black funnel naturally has a triangular integration domain with three boundaries: a horizon, the planar black hole metric, and the conformal boundary.  While it is possible to solve PDEs in triangular domains, the task is much more cumbersome, and it is more difficult to achieve the same accuracy as that of rectangular domains.  However, with an appropriate choice of conformal frame, this domain can be expanded to a square.  

By careful examination of the region where the bulk horizon meets the boundary horizon, one can show that the metric approaches that of a hyperbolic black hole\footnote{We thank Donald Marolf for this realization.}. In fact, this has to be the case for \emph{any} horizon that reaches the conformal boundary. Close to the conformal boundary, and up to terms that only change the holographic stress energy tensor, the geometry is hyperbolic. Furthermore, the most general cohomogeneity one line element that is static and manifestly exhibits hyperbolic symmetry for each hyperslice of constant time and radius is that of a hyperbolic black hole.  A specific hyperbolic black hole metric is then fixed by a choice of the bulk horizon temperature and the boundary horizon temperature.  In this manuscript, we are only considering bulk horizons with the same temperature as the corresponding boundary black hole. The only such black hole metric compatible with the desired symmetries is given by the zero energy hyperbolic black hole:
\be
ds^2_{\mathbb{H}}=\frac{L^2}{z^2}\left[-(1-z^2)dt^2+\frac{dz^2}{1-z^2}+d\eta^2+\sinh^2\eta\, d\Omega_{d-3}^2\right]\;.
\label{eq:hyperbolic}
\ee
Our solutions must approach the $\eta\rightarrow\infty$ limit of \eqref{eq:hyperbolic}.

Therefore, the aim is to find a solution that has a horizon, approaches the planar black hole and the hyperbolic black hole on either ends of this horizon, and has a boundary that is conformal to the background metric of the field theory. The above set of considerations changed the integration domain from a triangle into a square, see Fig.~\ref{fig:domain}.
\begin{figure}[t]
\centerline{
\includegraphics[width=0.8\textwidth]{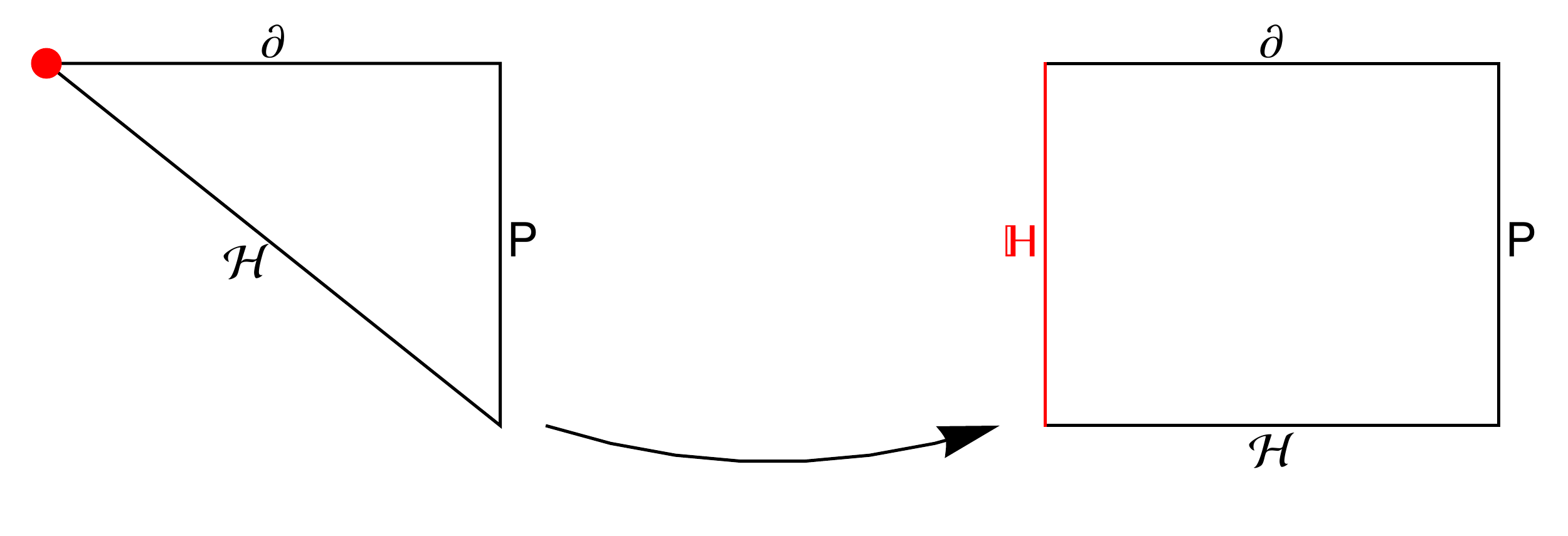}
}
\caption{Implicit coordinate transformation from a triangular to a rectangular domain integration domain. The point where the horizon meets the boundary is blown into a hyperslice where the black funnel line element approaches Eq.~(\ref{eq:hyperbolic}). In this diagram, $\mathbb{H}$ represents the hyperbolic black hole, $\partial$ the conformal boundary, $P$ the planar black hole and $\mathcal{H}$ the bulk horizon.}
\label{fig:domain}
\end{figure}
\end{subsection}

\begin{subsection}{A Schwarzschild Black Funnel}
For a Schwarzschild background, consider the following ansatz in $d=5$ dimensions:
\begin{multline}
ds^2=\frac{L^2}{x\,y(1+x)^2}\bigg\{-x(1-y)(1+x\,y)Tdt^2+\frac{x(1+x)^2Ady^2}{4y(1-y)(1+x\,y)}+
\\
+\frac{r_0^2B[dx+x(1-x)^2Fdy]^2}{x(1-x)^4}+\frac{r_0^2S}{(1-x)^2}d\Omega_{2}^2\bigg\}\;,
\label{5dansatz}
\end{multline}
where $T$, $A$, $B$, $F$, and $S$ are all functions of $x$ and $y$.  Also, $r_0=1/2$, a choice that we will explain shortly.  We choose the line element of our reference metric by setting $T=A=B=S=1$, and $F=0$.

Now let us discuss the boundary conditions.  At $x=1$, we would like our metric to approach that of the planar black hole.  This would mean that far from the boundary black hole in the field theory, the stress tensor becomes a thermal fluid.  Imposing Dirichlet boundary conditions, namely $T=A=B=S=1$, and $F=0$, the line element then reduces to:
\be
ds^2=\frac{L^2}{y}\left[-\frac{1}{4}(1-y^2)dt^2+\frac{dy^2}{4y(1-y^2)}+\frac{r_0^2dx^2}{4 (1-x)^4}+\frac{r_0^2}{4(1-x)^2}d\Omega_2^2\right]\;.
\ee
Under the coordinate transformation $t = 2 \tau $, $x = 1-r_0/2R$ and $ y= z^2$ the former line element becomes
\be
ds^2=\frac{L^2}{z^2}\left[-(1-z^4)d\tau^2+\frac{dz^2}{1-z^4}+dR^2+R^2d\Omega_2^2\right]\;,
\ee
which can be recognized as the line element of a five-dimensional planar black hole.

Our conformal boundary lies at $y=0$.  If we again choose $T=A=B=S=1$, and $F=0$, the metric becomes
\be
ds^2=\frac{L^2}{y}\left[\frac{dy^2}{4 y}+\frac{1}{x(1+x)^2}\;ds_\p^2\right]\;,
\ee
\be
ds_\p^2=-x\,dt^2+\frac{r_0^2dx^2}{x(1-x)^4}+\frac{r_0^2}{(1-x)^2}d\Omega_2^2\;,
\ee
where $ds_\p^2$ is the line element of our conformal boundary metric.  With the coordinate transformation $x= 1-r_0/r$, it becomes a more familiar form of the Schwarzschild metric:
\be
ds_\p^2=-\left(1-\frac{r_0}{r}\right)dt^2+\frac{dr^2}{1-\frac{r_0}{r}}+r^2d\Omega_2^2\;.
\label{eq:lineelmentschwarzschild}
\ee
Of course, the Schwarzschild black hole has the temperature
\be\label{bndrytemp}
T_\p=\frac{1}{4\pi r_0}\;.
\ee

The horizon of our funnel lies at $y=1$, where we impose regularity.  Note that at $y=1$, the $dt^2$ and $dy^2$ terms of the metric are the same if $T=A$.  This is necessary in order for the boundary conditions to be consistent with $\xi=0$, given our choice of reference metric.  Expanding the equations of motion about the horizon will give the condition $T=A$ as well as other conditions on $\p_y A|_{y=1}$, $\p_y B|_{y=1}$, $\p_y F|_{y=1}$, and $\p_y S|_{y=1}$.  

The temperature associated with the funnel horizon is given by
\be\label{tempschwarzschild}
T=\frac{1}{2\pi}\;.
\ee
We would like the temperature of our black funnel to match the temperature of the boundary black hole.  In the field theory, this means that the black hole must have the same temperature as its surrounding plasma, ensuring that we are constructing the gravitational dual to the Hartle-Hawking state. Therefore, \eqref{bndrytemp} and \eqref{tempschwarzschild} imply that $r_0=1/2$, justifying the choice made below \eqref{5dansatz}.

Finally, there is another boundary condition at $x=0$ where we would like our solution to approach the zero-energy hyperbolic black hole (\ref{eq:hyperbolic}).  We impose the same Dirichlet boundary condition $T=A=B=S=1$, and $F=0$, and the metric becomes:
\be
\label{x0metricschwarzschild}
ds^2=\frac{L^2}{y}\left[-(1-y)dt^2+\frac{dy^2}{4y(1-y)}+\frac{dx^2}{4x^2}+\frac{1}{4x}d\Omega_2^2\right]\;,
\ee
where we have substituted $r_0 = 1/2$.  Both Eq.~(\ref{x0metricschwarzschild}) and Eq.~(\ref{eq:hyperbolic}) should agree with each other close to $x\to0$ ($\eta\to+\infty$). In order to see this, consider the following change of variables $y=z^2$ and $x=\exp(-2\eta)$. Under this coordinate transformation (\ref{x0metricschwarzschild}) reduces to
\be
\label{x0metricnew}
ds^2=\frac{L^2}{z^2}\left[-(1-z^2)dt^2+\frac{dz^2}{1-z^2}+d\eta^2+\frac{e^{2\eta}}{4}d\Omega_2^2\right]\;,
\ee
which matches (\ref{eq:hyperbolic}) as $\eta\to+\infty$.

We can finally complete our proof for the non-existence of Ricci solitons. In order to do this, we need to ensure that $\chi$ is zero on $\mathbb{H}$, $\partial$, $\mathcal{H}$ and $P$, as in Fig.~\ref{fig:domain}. This was already shown in \cite{Figueras:2011va} for the first three boundaries. So, we only need to show that this is the case for our new boundary $P$. It turns out that with our choice of reference metric, $\chi = \mathcal{O}[(1-x)^2]$, thus ensuring that no Ricci solitons exist.

\end{subsection}
\begin{subsection}{Black Funnels with $d=4$}
Now let's consider the $d=4$ ansatz:
\begin{align}\label{4dansatz}
ds^2=\frac{L^2}{x^2y^2g(x)^2}\bigg[&-x^2(1-y)f(x,y)Tdt^2+\frac{x^2g(x)^2Ady^2}{(1-y^2)f(x,y)}+\nonumber\\
&\qquad+\frac{4B(dx+x(1-x^2)^2Fdy)^2}{(1-x^2)^4}+\frac{\ell(x)S}{(1-x^2)^2}d\phi^2\bigg]\;,
\end{align}
with
\be\label{fgell}
f(x,y)=1+y+x^2y^2(3-2x)\,\qquad g(x)=2+x^2(3-2x)\,\qquad \ell(x)=1+\lambda(1-x^2)\;.
\ee
As before, $T$, $A$, $B$, $F$, and $S$ are all functions of $x$ and $y$.  The reference metric again has $T=A=B=S=1$, and $F=0$.  The constant $\lambda$ controls the size of the boundary black hole.  The factors of $3-2x$ were added to improve numerics by reducing steep gradients.  

The boundary conditions are essentially the same as the Schwarzschild funnel.  At $x=1$, we require $T=A=B=S=1$, and $F=0$ and recover the planar black hole:
\be
ds^2=\frac{L^2}{y^2}\left[-\frac{1}{9}(1-y^3)dt^2+\frac{dy^2}{1-y^3}+\frac{4dx^2}{9(1-x^2)^4}+\frac{d\phi^2}{9(1-x^2)^2}\right]\;,
\ee
This can be put in a more familiar form by the transformation $t=3\tau$, $x=\sqrt{1-1/3R}$, which at large $R$ becomes
\be
ds^2=\frac{L^2}{y^2}\left[-(1-y^3)d\tau^2+\frac{dy^2}{1-y^3}+dR^2+R^2d\phi^2\right]\;.
\ee
At the conformal boundary $y=0$,  we again choose $T=A=B=S=1$, and $F=0$:
\be
ds^2=\frac{L^2}{y^2}\left[dy^2+\frac{1}{x^2g(x)^2}\;ds_\p^2\right]
\ee
\be\label{dsp2}
ds_\p^2=x^2dt^2+\frac{4dx^2}{(1-x^2)^4}+\frac{1+\lambda(1-x^2)}{(1-x^2)^2}d\phi^2\;,
\ee
With the coordinate transformation $x\rightarrow \sqrt{1-1/r}$, our boundary metric becomes
\be\label{bndrymetric2}
ds_\p^2=-\left(1-\frac{1}{r}\right)dt^2+\frac{dr^2}{1-\frac{1}{r}}+r^2\left(1+\frac{\lambda}{r}\right)d\phi^2\;.
\ee
This is the metric of an asymptotically flat three dimensional black hole with horizon at $r=1$.  The size of the black hole is given by the size of the circle at the horizon: $\sqrt{1+\lambda}$.  This black hole is not Ricci flat and does not solve the vacuum Einstein equation.  Here, the boundary black hole is merely a boundary condition and does not need to satisfy any equations of motion.  In the field theory, this amounts to choosing a fixed background. 

At $y=1$, we have a horizon where we require regularity.  As before, the $dt^2$ and $dy^2$ terms are the same if $T=A$.  The temperature of this black hole is given by
\be\label{temp}
T=\frac{1}{4\pi}\;,
\ee
which is the same temperature as our boundary black hole.  

At $x=0$, we again choose the same boundary condition $T=A=B=S=1$, and $F=0$, and the metric becomes a four-dimensional zero energy hyperbolic black hole \eqref{eq:hyperbolic} after the coordinate redefinition $t=2\tau$, $y=z$, and $x = \sqrt{1+\lambda}\,\exp(-\eta)$:
\be\label{x0metric}
ds^2=\frac{L^2}{z^2}\left[-(1-z^2)d\tau^2+\frac{dz^2}{1-z^2}+d\eta^2+\frac{e^{2\eta}}{4}d\phi^2\right]\;.
\ee

As in the Schwarzchild case, to show non-existence of Ricci solitons, all that remains is to show that $\chi$ vanishes on the boundary at $x=1$.  We find that $\chi = \mathcal{O}[(1-x)^2]$ so no solitons exist in this case either.  
\end{subsection}
\begin{subsection}{Numerics}
To solve the PDEs numerically, we approximate the system by a set of non-linear algebraic equations using pseudospectral collocation on a Chebyshev grid.  We then solve the algebraic equations with a standard Newton-Raphson relaxation procedure.  This method requires a seed, which we choose to be the reference metric.  

This method has the advantage that it achieves exponential convergence as the number of grid points is increased.  We only require a few points (around $40\times40$), and modest computational resources.  However, a drawback is that the solutions must be analytic.  The presence of logarithms from a conformal anomaly prevent us from varying the size of the black funnel in five dimensions.  

As mentioned earlier in this section, Ricci solitons have been proven not to exist for our system. We therefore use the norm of the DeTurck vector $\chi \equiv \xi_\mu\xi^\mu$ to monitor our numerical error.  In Fig.~\ref{fig:convergence}, we plot the maximum value of $\chi$ as a function of grid points and show the expected exponential convergence.
\begin{figure}[t]
\centerline{
\includegraphics[width=0.45\textwidth]{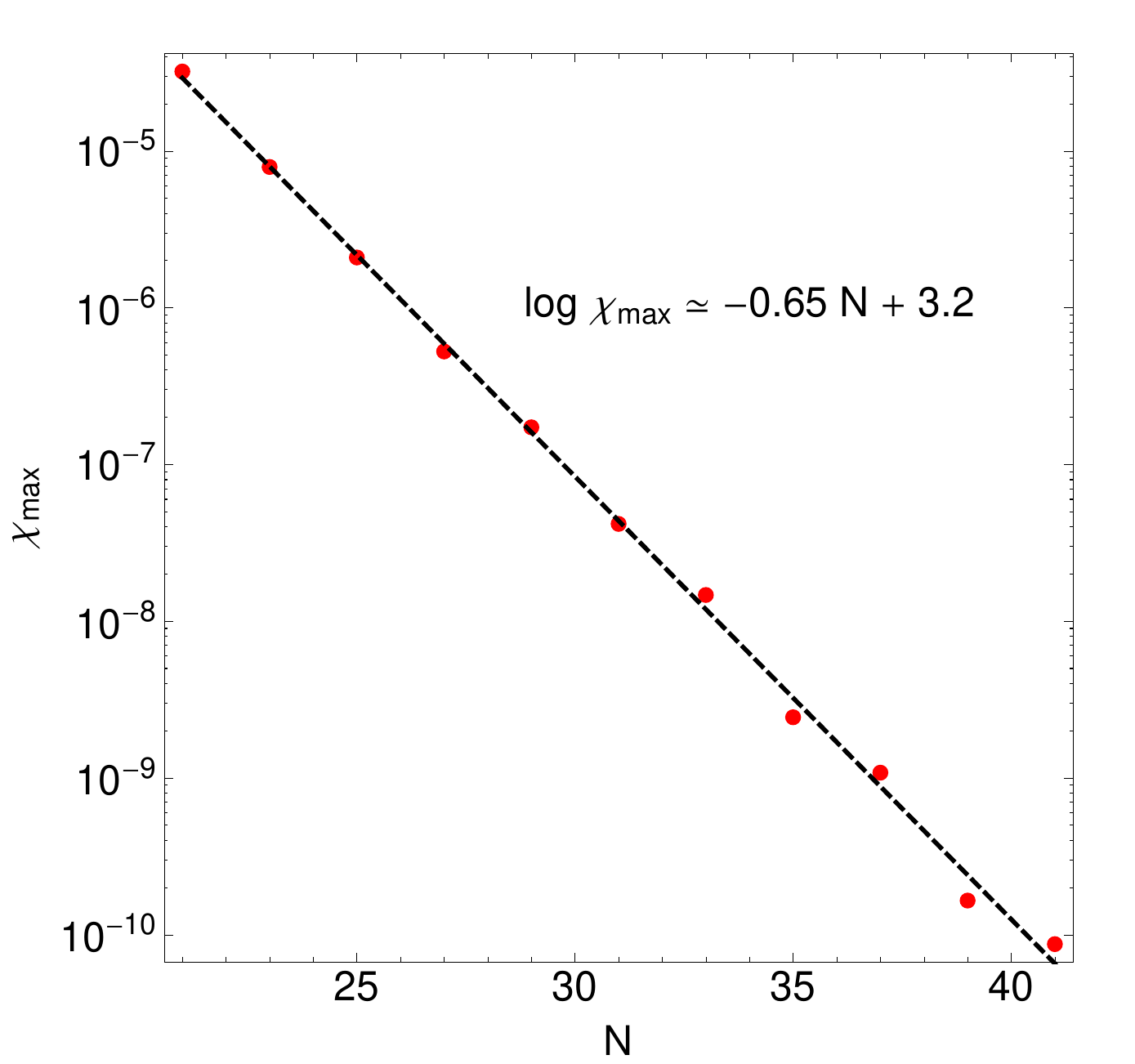}
\hspace{1cm}
\includegraphics[width=0.43\textwidth]{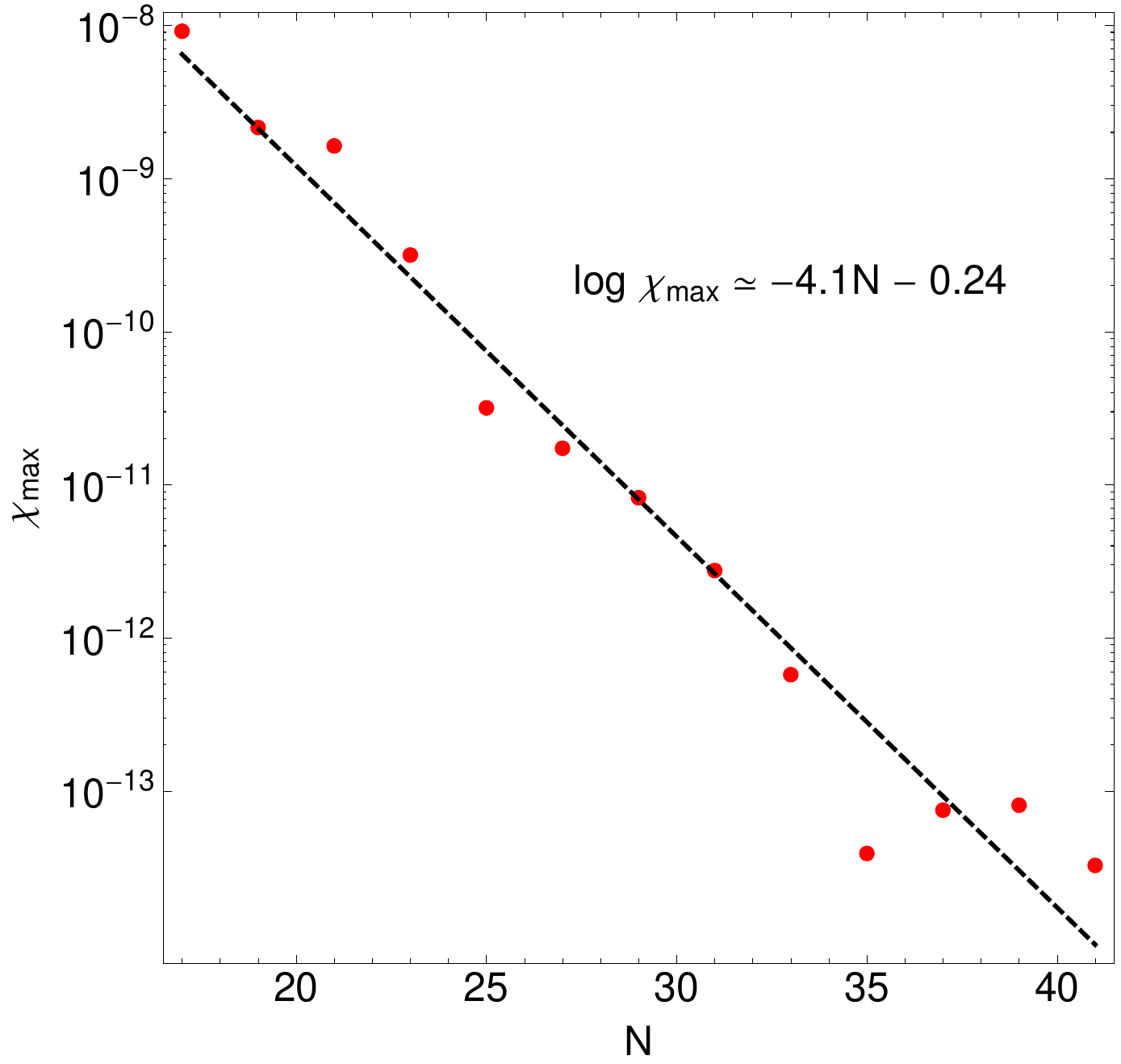}
}
\caption{The maximum value of $\chi$ as a function of the number of grid points $N$ for the Schwarszchild funnel (left) and the four-dimensional funnel with $\lambda = 0$ (right). The vertical scale is logarithmic, and the data is well fit by an exponential decay.  There is some noise in the convergence below $10^{-11}$ from machine precision.}
\label{fig:convergence}
\end{figure}
\end{subsection}
\end{section}
\begin{section}{Boundary Stress Tensor}
\begin{subsection}{Schwarzschild Funnel Stress Tensor}
We now compute the stress tensor for our solutions.  We will closely follow de Haro, Skenderis and Solodukhin in \cite{de Haro:2000xn}. We will start by analyzing the five-dimensional funnel.  As a first step, we need to solve the field equations in a series expansion off the conformal boundary, located at $y=0$. In order to compute the holographic stress energy tensor in $d=5$, we only need to solve Eqs.~(\ref{deturck}) to $\mathcal{O}(y^3)$. Up to this order, we find the following expansions
\begin{subequations}
\begin{multline}
T(x,y) = 1+\frac{1}{2} (1-x) x [(1-x) x (3 x^2-8 x+6)+1]\, y\\
-\frac{(1-x) x [x L_T(x)-4 (1+x) b_2^{\prime}(x)]+(6+4 x-34 x^2) b_2(x)}{2 (1-3 x) (1+x)}\,y^2+\mathcal{O}(y^3)\,,
\end{multline}
\begin{equation}
A(x,y) = 1-\frac{1}{4} (1-x)^2 x L_A(x)\,y^2+\mathcal{O}(y^3)\,,
\end{equation}
\begin{equation}
F(x,y) = (1 - x)^3 x^2 (1 + x) L_F(x)\,y+\mathcal{O}(y^2)\,,
\end{equation}
\begin{equation}
B(x,y) = 1-\frac{x}{2}(27 x^5-46 x^4-23 x^3+76 x^2-35 x+1)\,y+b_2(x)\,y^2+\mathcal{O}(y^3)\,,
\end{equation}
\begin{multline}
S(x,y) = 1-\frac{1}{2} (1-x) x [(1-x) x (14+8 x-21 x^2)+1]\, y\\
+\frac{4 (1+2x-7x^2) b_2(x)-(1-x) x [1+4 (1+x) b_2^{\prime}(x)-x L_S(x)+]}{4 (1-3 x) (1+x)}\,y^2+\mathcal{O}(y^3)\,,
\end{multline}
\label{eqs:asymptoticexpansion}
\end{subequations}
where $L_{T}$, $L_A(x)$, $L_F(x)$ and $L_S(x)$ are polynomials which are given in Appendix \ref{appendix:a}. This expansion is compatible with $\xi = 0$ up to $\mathcal{O}(y^3)$. The only free function in this expansion is $b_2(x)$, which can be extracted by taking two derivatives of $B$ with respect to $y$. In this way, we have no need to resort to a fitting procedure, thus increasing our numerical accuracy in determining the holographic stress energy tensor.

The next step is to change to the familiar Fefferman-Graham coordinates, in which case the metric can be recast as
\begin{equation}
ds^2 = \frac{L^2}{\tilde{z}^2}\left[d\tilde{z}^2+ds^2_\p+\tilde{z}^2 h_2+\tilde{z}^4 h_4+\mathcal{O}(\tilde{z}^6)\right]\,,
\end{equation}
where $ds^2_\p$ is defined in \eqref{eq:lineelmentschwarzschild}. This coordinate transformation is performed in an expansion in $\tilde{z}$, where the successive terms are determined by requiring $g_{\tilde{z} \mu}=0$. The coordinate transformation has the following schematic form:
\begin{align}
&y = \frac{\tilde{z}^2}{\tilde{x}(1+\tilde{x})^2}+R_4(\tilde{x})\tilde{z}^4+R_6(\tilde{x})\tilde{z}^6+\mathcal{O}(\tilde{z}^8)\nonumber
\\
\label{eq:tofeffer}
\\
&x = \tilde{x}+\frac{(1-\tilde{x})^4(1+3\tilde{x})}{1+\tilde{x}}\tilde{z}^2+V_4(\tilde{x})\tilde{z}^4+\mathcal{O}(\tilde{z}^6)\nonumber
\end{align}
where the remaining $R_i$ and $V_i$ are given in Appendix \ref{appendix:a}. The $\tilde{z}^2$ term in the expansion of $y$ fixes the conformal frame we are interested in, namely one where the boundary metric has $\tilde{g}_{tt} = -\tilde{x}$.

After reading off $h_4$, we can reconstruct the corresponding holographic stress energy tensor \cite{de Haro:2000xn} via
\begin{equation}
\langle T_{ij}\rangle = \frac{h_4}{4\pi G_5}\;.
\end{equation}
where $i$ and $j$ now run over the coordinates in the boundary metric.  Note that because our conformal boundary metric is Ricci flat, there is no conformal anomaly and $h_2$ is identically zero. The holographic stress energy tensor is then given by:
\begin{multline}
4\pi G_5\langle T_{ij}\rangle = \mathrm{diag}\bigg\{\frac{\tilde{x} [\tilde{x} \tilde{T}_{tt}(\tilde{x})-16 (1-\tilde{x}^2) b_2^{\prime}(\tilde{x})]+8 \left(3+2 \tilde{x}-17 \tilde{x}^2\right)b_2(\tilde{x})}{8 (1-3 \tilde{x}) \tilde{x} (1+\tilde{x})^5},\\
-\frac{\tilde{x}^2 \tilde{T}_{\tilde{x}\tilde{x}}(\tilde{x})-8 b_2(\tilde{x})}{32 \tilde{x}^3 (1-\tilde{x}^2)^4},\frac{\tilde{x} [\tilde{x} \tilde{T}_{\Omega}(\tilde{x})-8 (1-\tilde{x}^2) b_2^{\prime}(\tilde{x})]+8 (1+2 \tilde{x}-7 \tilde{x}^2)b_2(\tilde{x})}{32 (1-3 \tilde{x}) \tilde{x}^2 (1+\tilde{x})^3 (1-\tilde{x}^2)^2}\bigg\}\,.
\label{eq:holostressschwarzschild}
\end{multline}
This stress energy tensor is manifestly traceless and transverse with respect to (\ref{eq:lineelmentschwarzschild}), as expected for a conformal field theory on a curved Ricci flat background. $\tilde{T}_{ij}$ are polynomials whose expressions are given in Appendix \ref{appendix:a}.

Finally, in Fig.~\ref{fig:stressenergytensorshwarzschild} we show how $\langle T_t^{\;t}\rangle$ varies with $r$, where $r$ is the radial direction in Schwarzschild coordinates defined in Eq.~(\ref{eq:lineelmentschwarzschild}). The dashed line represents the value of $\langle T_t^{\;t}\rangle$ evaluated in the planar black hole geometry.  It is reassuring that the numerical data (represented by points) approaches the dashed line for large $r$. For $r/r_0>10$, we fit this component of the stress energy tensor to a function of the form
\begin{equation}
P_{d=5}(r)=\frac{A_0}{r^\alpha}\left(1+\frac{B_0}{r}\right)-\frac{3}{256 \pi},
\end{equation}
where the last term corresponds to the value of $\langle T_t^{\;t}\rangle$ for the planar black hole. A $\chi^2$ fit yields $A_0 \approx -0.007$, $B_0 \approx 2.181$ and $\alpha \approx 0.986$, which is consistent with a $1/r$ tail in the stress energy tensor. This should be contrasted with the faster fall off for the same component of the stress energy tensor found in \cite{Figueras:2011va}.  The slower falloff for the funnels is indicative of the stronger coupling between the boundary black hole and its surrounding plasma.  Finally, we note that $\langle T_t^{\;t}\rangle$ is smooth across the boundary black hole horizon, as expected.

\begin{figure}[t]
\centerline{
\includegraphics[width=0.5\textwidth]{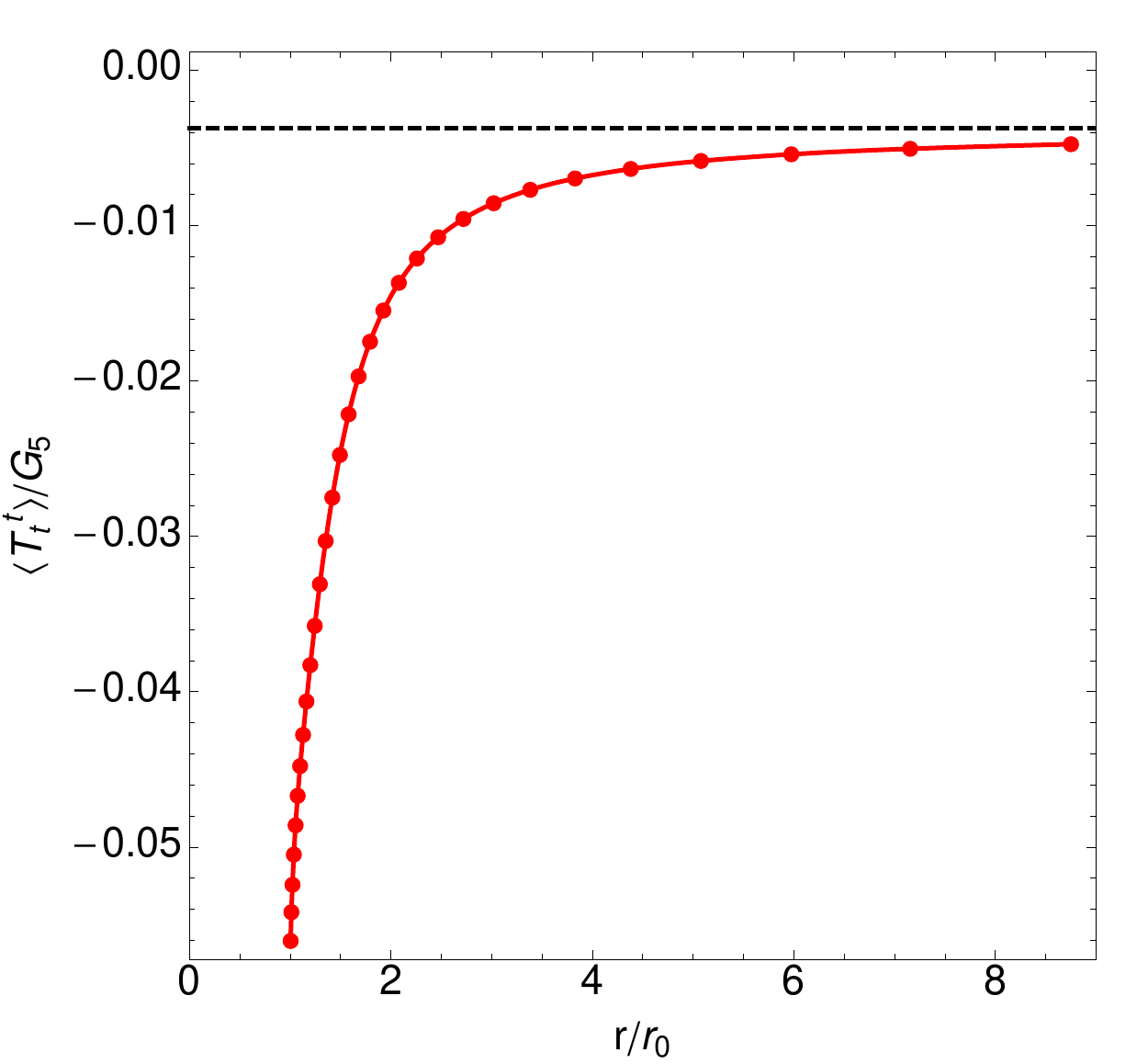}
}
\caption{Plot of $\langle T_t^{\;t}\rangle$ as a function of $r/r_0$. The dashed line indicates the value of $\langle T_t^{\;t}\rangle $ computed for the planar black hole, while the red dots represent our numerical data extracted after solving (\ref{deturck}) with appropriate boundary conditions. This particular run was done with 41 points along each Chebyshev grid.}
\label{fig:stressenergytensorshwarzschild}
\end{figure}
\end{subsection}

\begin{subsection}{$d=4$ Funnel Stress Tensor}
Now we repeat this construction for the four-dimensional funnels.  An expansion of the equations of motion off the conformal boundary gives
\begin{subequations}
\begin{equation}
T(x,y) = 1+\frac{1}{8}(1-x)^2K_T(x)\, y^2+t_3(x)\,y^3+\mathcal{O}(y^4)\,,
\end{equation}
\begin{equation}
A(x,y) = 1+\mathcal{O}(y^4)\,,
\end{equation}
\begin{equation}
F(x,y) = \frac{1}{16}(1 - x) x^2 (1 - x^2)^2g(x) K_F(x)\,y^3+\mathcal{O}(y^4)\,,
\end{equation}
\begin{equation}
B(x,y) = 1+\frac{1}{8}(1-x)^2K_B(x)\,y^2+b_3(x)\,y^3+\mathcal{O}(y^4)\,,
\end{equation}
\begin{equation}
S(x,y) =1+\frac{1}{8}(1-x)^2K_S(x)\, y^2+s_3(x)\,y^3+\mathcal{O}(y^4)\,,
\end{equation}
\label{eqs:asymptoticexpansion2}
\end{subequations}
where $K_T$, $K_F$, $K_B$, and $K_S$ are known functions defined in Appendix A.   Furthermore, $t_3$ and $s_3$ can be expressed in terms of $b_3$ via
\begin{subequations}\label{tbs3}
\begin{align}
t_3(x)&=-s_3(x)-b_3(x)\;,\\
s_3(x)&=b_3(x)-\frac{x(1-x^2)\ell(x)\left[4(1-x)^3x(1+2x)-18(1-x)x b_3(x)+g(x)b_3'(x)\right]}{g(x)\left[x^2(1-x^2)\lambda+(1-3x^2)\ell(x)\right]}\;,
\end{align}
\end{subequations}
where we have defined $\ell$ and $g$ in \eqref{fgell}.  

Now we go to Fefferman-Graham coordinates:
\begin{equation}
ds^2 = \frac{L^2}{\tilde{z}^2}\left[d\tilde{z}^2+ds^2_\p+\tilde{z}^2 h_2+\tilde{z}^3 h_3+\mathcal{O}(\tilde{z}^6)\right]\,
\end{equation}
where $ds^2_\p$ is given by \eqref{dsp2}. The coordinate transformation that does this is
\begin{align}
&y = \frac{\tilde{z}}{\tilde{x}g(\tilde x)}+Q_3(\tilde x)\tilde{z}^3-\frac{3-2\tilde x}{6\tilde x^2g(\tilde x)^4}z^4+\mathcal{O}(\tilde{z}^6)\nonumber
\\
\label{eq:tofeffer2}
\\
&x = \tilde{x}+\frac{(1-\tilde{x}^2)^4[2+\tilde{x}^2(9-8\tilde x)]}{8\tilde xg(\tilde x)}\tilde{z}^2+U_4(\tilde{x})\tilde{z}^4+\mathcal{O}(\tilde{z}^6)\nonumber\;,
\end{align}
where $Q_3$ and $U_4$ are given in Appendix A.  The holographic stress energy tensor is then given by \cite{de Haro:2000xn}
\begin{equation}
\langle T_{ij}\rangle = \frac{3h_3}{16\pi G_4}\;.
\end{equation}
Although the four-dimensional funnels are not Ricci flat, there is no conformal anomaly in this dimension.  Then the boundary stress tensor is
\begin{multline}
\frac{16\pi G_4}{3}\langle T_{ij}\rangle = \mathrm{diag}\bigg\{\frac{2g(\tilde x)-3 t_3(\tilde x)-4}{3\tilde x g(\tilde x)^3},\frac{4(g(\tilde x)+3 b_3(\tilde x)-2)}{3\tilde x^3(1-\tilde x^4)^4g(\tilde x)^3},\frac{\ell(\tilde x)[g(\tilde x)+3s_3(\tilde x)-2]}{3\tilde x^3(1-\tilde x^2)^2g(\tilde x)^3}\bigg\}\;.
\label{eq:holostressschwarzschild2}
\end{multline}
Combined with \eqref{tbs3}, this stress tensor is manifestly transverse and traceless.  

In Fig.~\ref{fig:stressenergytensorshwarzschild2}, we plot $\langle T_t^{\phantom{t}t}\rangle$ as a function of the radial coordinate $r$ defined in \eqref{bndrymetric2}, for various values of $\lambda$. The behaviour at large $r$ also approaches that of the planar black hole, which has the value shown by the dashed line.  As before, we can fit this component of the stress energy tensor for all points $r>10$ to 
\begin{equation}\label{fitfunction2}
P_{d=4}(r)=\frac{A_0}{r^\alpha}\left(1+\frac{B_0}{r}\right)-\frac{1}{216 \pi}\;.
\end{equation}
We find that for all values of $\lambda$ between $0$ and $-0.9$, $A_0\approx0.0021$, $B_0\approx1.8$, and $\alpha\approx0.99$.  The four-dimensional funnels give the same $1/r$ falloff as those in five dimensions.  This is expected since both boundary metrics approach Minkowski space asymptotically as $1/r$. 

We also see that $\langle T_t^{\phantom{t}t}\rangle $ is smooth across the boundary black hole horizon. In addition, as the size of the black hole is decreased, the value of $\langle T_t^{\phantom{t}t}\rangle$ increases at the horizon.  Moreover, for small black holes, \emph{i.e.} $\lambda \sim -1$, $\langle T_t^{\phantom{t}t}\rangle$ at the horizon behaves as $\mathcal{R}^{3/2}$, where $\mathcal{R}$ represents some measure of the curvature, such as the square root of the Kretschmann scalar at the horizon of the three-dimensional boundary black hole.

\begin{figure}[t]
\centerline{
\includegraphics[width=0.5\textwidth]{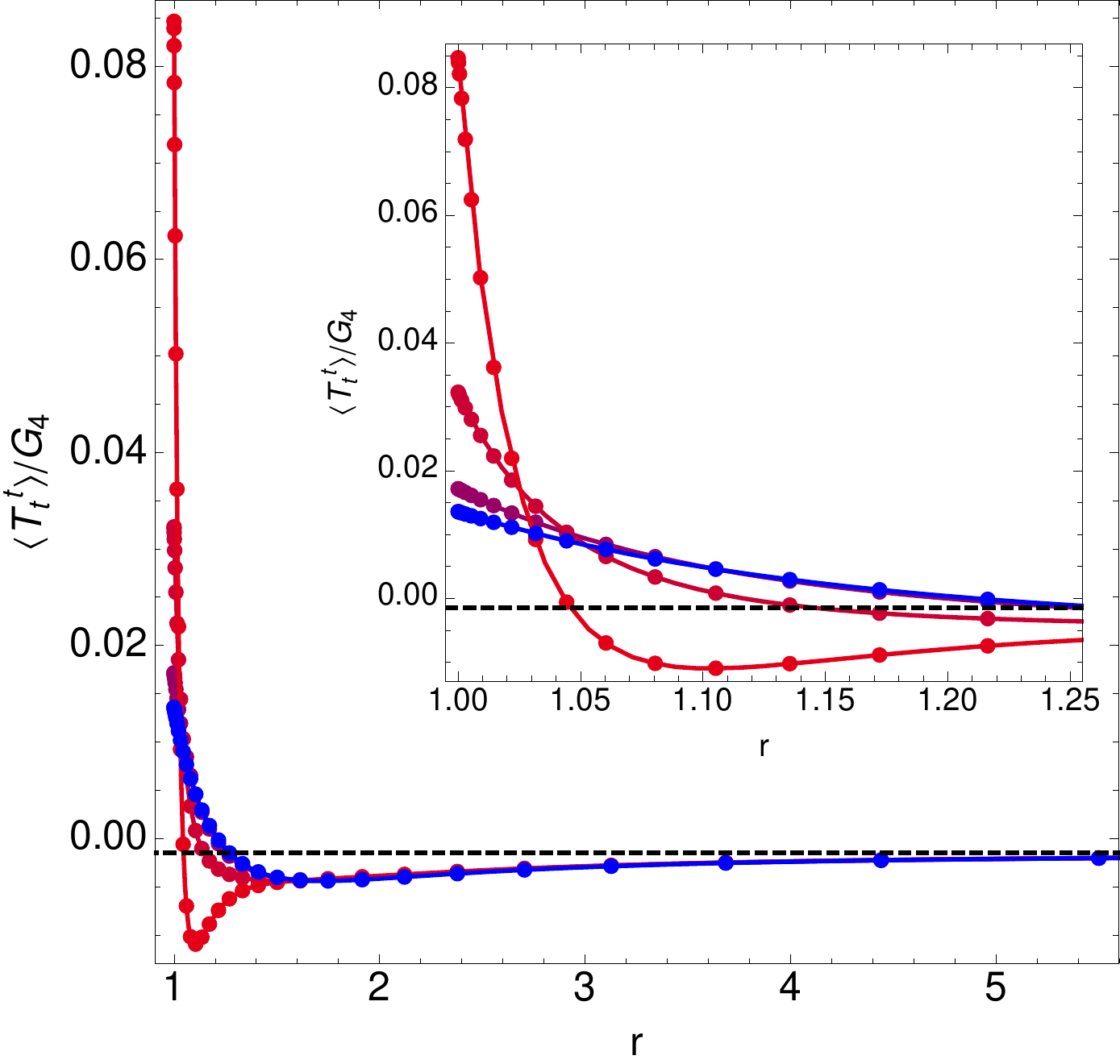}
}
\caption{Plot of $\langle T_t^{\;t}\rangle$ as a function of $r$ for the four-dimensional funnel with $\lambda$ taking values of $0$, $-0.6$, $-0.8$, and $-0.9$. Lowering $\lambda$ increases $\langle T_t^{\;t}\rangle$ at the horizon $r=1$.  The dashed line indicates the value of $\langle T_t^{\;t}\rangle$ computed for the planar black hole, while the coloured dots represent our numerical data.  We used a $40\times40$ Chebyshev grid.}
\label{fig:stressenergytensorshwarzschild2}
\end{figure}
\end{subsection}
\end{section}

\begin{section}{Embedding into Hyperbolic Space}
In this section, we will visualize the horizon of our funnels by embedding them in hyperbolic space. Regardless of the dimension, the induce metric on the spatial horizon of our funnels has the same schematic form:
\begin{equation}
ds^2=\frac{L^2}{(1-x)^2}\left[\frac{\tilde{g}^{(d)}_{xx}(x)\,dx^2}{(1-x)^2}+\tilde{g}^{(d)}_{\Omega}(x)d\Omega^2_{d-3}\right]\,,
\label{eq:inducedmetricH}
\end{equation}
where $\tilde{g}^{(d)}_{xx}(x)$ and $\tilde{g}^{(d)}_{\Omega}(x)$ are functions that depend only on $x$ and can be read from the pullback of the line elements (\ref{5dansatz}) and (\ref{4dansatz}) to the $y=1$ hyperslice.

On the other hand, $d-1$ Euclidean hyperbolic space has the following form
\begin{equation}
ds^2_{\tilde{\mathbb{H}}_{d-1}}=\frac{L^2}{\tilde{y}^2}(d\tilde{y}^2+dR^2+R^2 d\Omega^2_{d-3})\,.
\label{eq:hyperbolicdm1}
\end{equation}
An isometric embedding can be easily achieved if we consider a curve of the form $s(x)=(\tilde{y}(x),R(x))$ in hyperbolic space. The pullback of the line element $(\ref{eq:hyperbolicdm1})$ to $s$ gives the following $d-2$ induced metric:
\begin{equation}
ds^{2}_{s,\tilde{\mathbb{H}}_{d-1}}=\frac{L^2}{\tilde{y}^2(x)}\left[\left(\tilde{y}'(x)^2+R'(x)^2\right)dx^2+R^2(x) d\Omega^2_{d-3}\right]\,.
\label{eq:inducedhyper}
\end{equation}

Demanding that (\ref{eq:inducedhyper}) agrees with (\ref{eq:inducedmetricH}) gives a systems of two equations in $R(x)$ and $\tilde{y}(x)$. One of the equations is a simple algebraic equation that determines $R(x)$ as a function of $\tilde{y}(x)$, whereas the remaining equation is a first order non-linear ODE in $\tilde{y}(x)$ which we solve numerically on a Chebyshev grid. As a boundary condition, we fix $R(0)$ to be the size of the boundary horizon in the black hole frame. Note that hyperbolic space is appropriate for the visualization of our funnels since a Schwarzschild string in AdS would be represented by a line of constant $R$, and the planar black hole would be given by a line of constant $\tilde{y}$.

The embeddings are displayed in Fig.~\ref{figs:embeddings}. The left panel shows the Schwarzschild funnel while the right panel shows the four-dimensional ones. The vertical dashed line in Fig.~\ref{fig:subfig1} represents an embedding digram for a Schwarzschild string  with $r_0 = 1/2$. One can attempt to give an heuristic argument based on these embeddings in favor of the stability of the Schwarzschild funnel. The stability of the Schwarzschild string in AdS was first studied in \cite{Gregory:2000gf} where it was found that the string is Gregory-Laflamme unstable only when it exceeds a length of about $\Delta \tilde{y}\approx 2.94$. One can ask when our geometry significantly deviates from the AdS-Schwarzschild string\footnote{Note that a string is unstable if the mode that sits at the onset of the Gregory-Laflamme instability fits inside the string.}. In order to quantify this, we define $\widetilde{\Delta R} \equiv |1-R(x)/r_0|$. We do not consider the black funnel to be string-like if we find deviations above the five percent level, or equivalently when $\widetilde{\Delta R}<0.05$. Given that this occurs at about $\tilde{y}\simeq0.53$, we find it unlikely that this geometry will be Gregory-Laflamme unstable, since the unstable mode does not seem to fit the string-like region of the Schwarzschild funnel\footnote{We thank Donald Marolf for this heuristic argument.}. Indeed, we have attempted to directly study the stability of these objects to axisymmetric perturbations and our preliminary results indicate stability in this sector of the perturbations. We have also searched for negative modes and found none. In Fig.~\ref{fig:subfig2} one sees that the four-dimensional case is even less likely to be unstable, given that the four-dimensional funnels appear to widen more quickly than the five-dimensional case.  Even if the size is decreased, the four-dimensional funnels do not become more narrow.  Furthermore, there is no known Gregory-Faflamme instability in four dimensions.  
\begin{figure}[t]
	\centering
	\subfigure[$d=5$]
		{
			\includegraphics[width=0.45\textwidth]{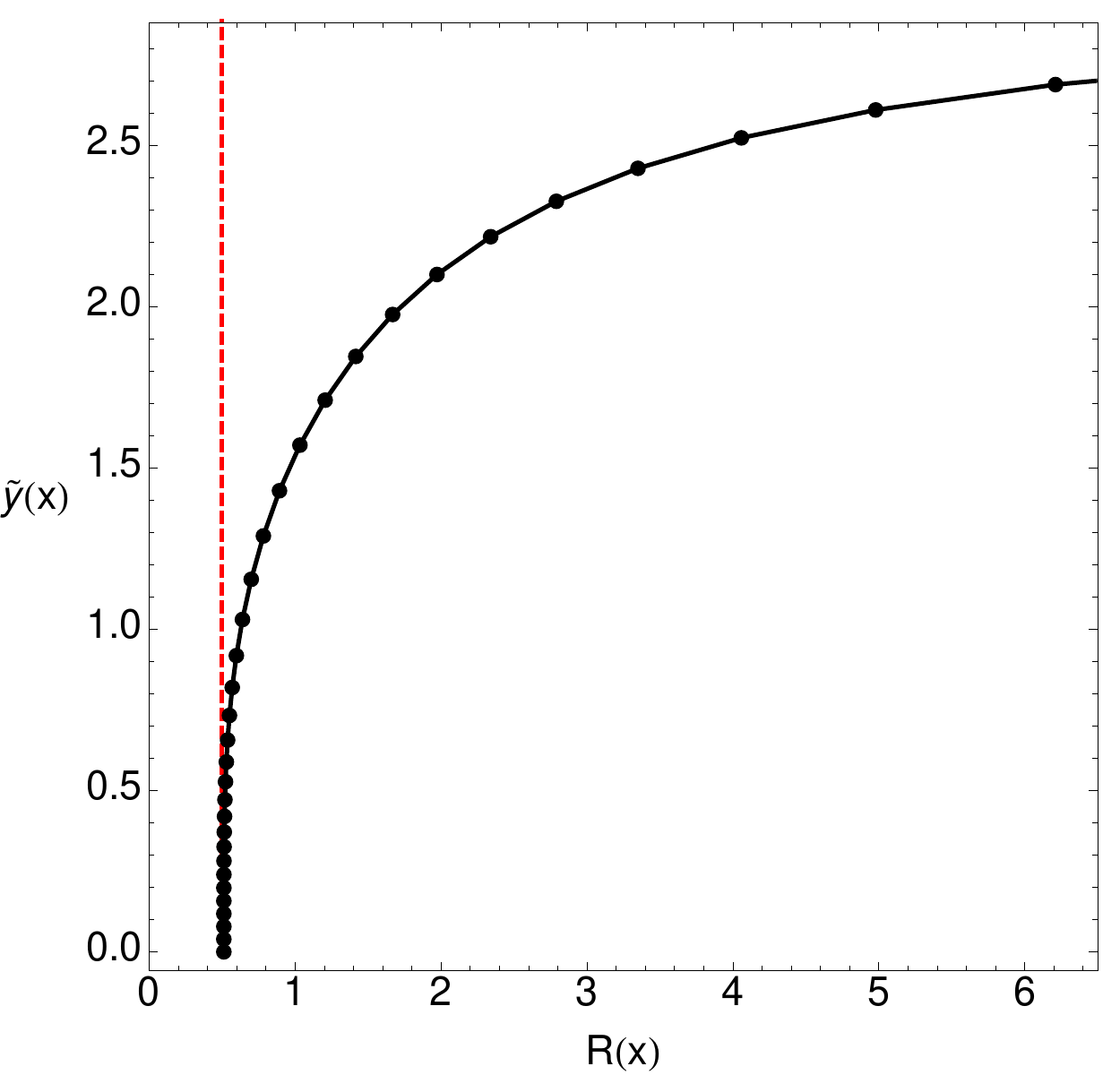}
			\label{fig:subfig1}
		}
	\subfigure[$d=4$]
		{
			\includegraphics[width=0.45\textwidth]{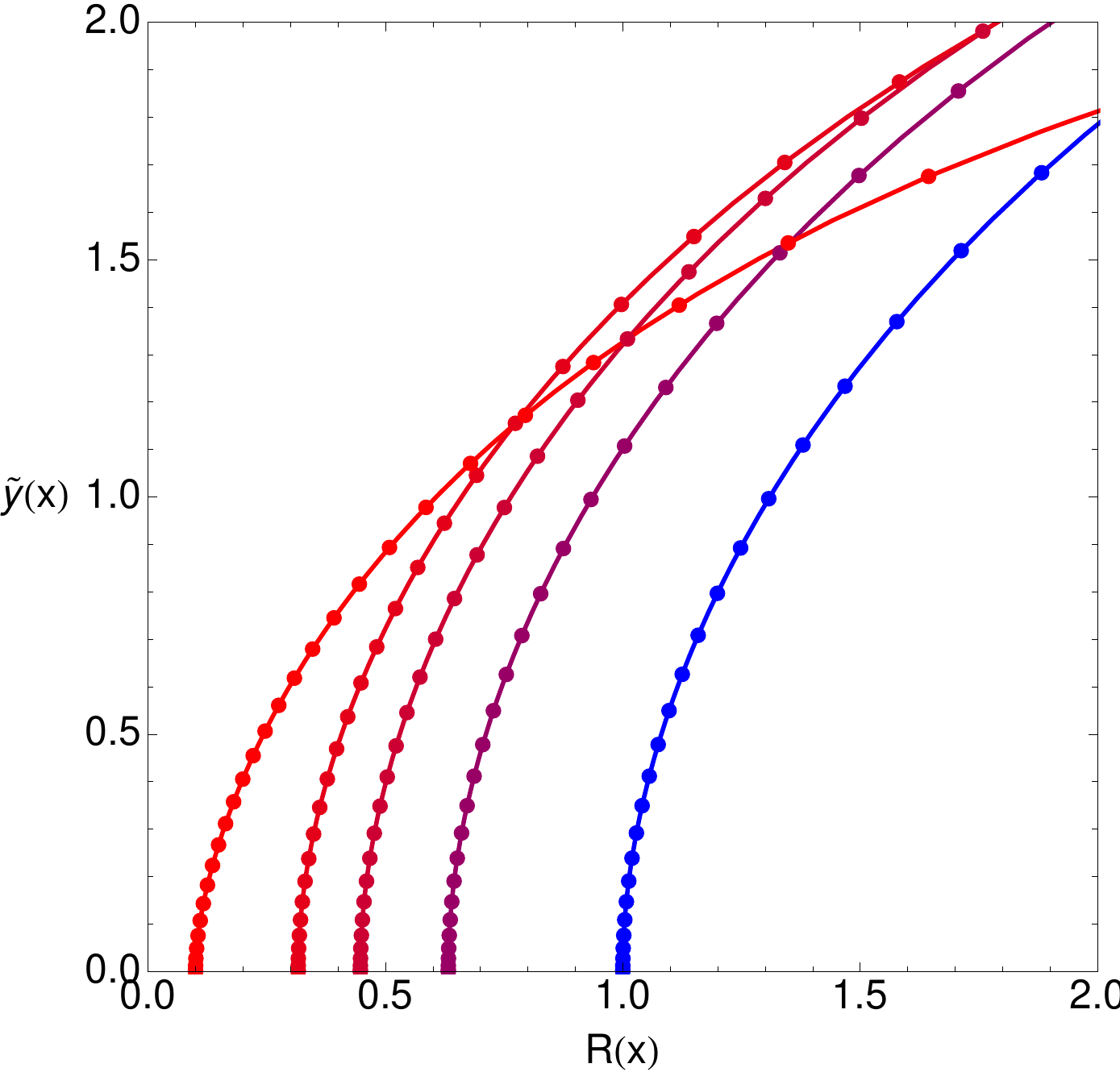}
			\label{fig:subfig2}
		}
	\caption{\label{figs:embeddings}Embedding diagrams of the spatial horizon induced metric for $d=4$ (left) and $d=5$ (right).  The dashed line represents the embedding of a Schwarzschild black string horizon.  The values for $\lambda$ in the $d=4$ plot are $0$, $-0.6$, $-0.8$, $-0.9$, and $-0.99$.}
\end{figure}

\end{section}

\begin{section}{Discussion} 
We have numerically constructed black funnel phases dual to strongly coupled Hartle-Hawking states on a Schwarzschild background and on a class of three dimensional black hole backgrounds.  We computed their holographic stress energy tensors and also varied the size of the three dimensional black hole backgrounds.  We also embedded the horizon geometry of these funnels in hyperbolic space, and argued that no Gregory-Laflamme instability occurs for these solutions.  

There are a number of future directions. First, it would be interesting to complete the stability analysis of the black funnels, \emph{i.e.} consider non-axisymmetric perturbations. Even though we don't envisage any other type of instability, a rigorous analysis would be most welcome. 

Here, we have only varied the size of the black hole in four dimensions where there is no known Gregory-Laflamme instability.  It would be interesting to see if such an instability occurs in higher dimensions.  One could attempt to find six-dimensional funnels where there is no conformal anomaly, or to vary the size of five-dimensional funnels using some other numerical method like finite differences. 

In order to fully test the phase transition conjectured in \cite{Hubeny:2009ru}, the black droplet phases must also be constructed.  These solutions naturally lie in a domain of integration with five boundaries: two horizons, an axis, the planar black hole metric, and the boundary metric.  This type of problem likely requires patching of separate integration domains, or some means of collapsing one of these boundaries to a point.  Three of these boundaries are also `fictitious' and require Neumann data rather than Dirichlet, making it more difficult to find a suitable seed.  Perhaps one can attempt to find solutions sufficiently close to the droplet in \cite{Figueras:2011va}, or obtain a suitable seed from a matched asymptotic expansion.   

Our solutions have bulk horizons that are at the same temperature as the boundary horizons.  One can imagine the field theory scenario where the boundary black hole is kept at a different temperature than its surrounding plasma.  In order to construct the bulk dual to this situation, one can use a different hyperobolic black hole than \eqref{eq:hyperbolic}.

One could attempt to find equivalent solutions in global AdS, as opposed to our funnels which can be thought of as lying in the Poincar\'{e} patch.  If the boundary field theory lies on a sphere with two black holes at the poles, there are now potentially three competing phases for the gravitational dual.  There is a funnel phase where the two boundary horizons are connected in the bulk.  But there are also two droplet phases where the boundary horizons are disconnected: one with a third spheroidal horizon in the bulk, and one without.  

Aside from Hawking radiation, black funnels have also been used to study heat transport \cite{Fischetti:2012ps}.  A black hole is viewed as a heat source in the dual field theory and the heat transport properties are mapped to a stationary flow in the black funnel.  One might be able to construct stationary black funnel solutions in order to study non-equilibrium heat flow.  This can also be attempted in global AdS.  

One can also imagine having a boundary black hole that rotates such as the Kerr metric.  Then the dual black funnels will also be rotating.  In order to prevent superluminal motion far from the axis of rotation, the horizons must `twist'.  These solutions will therefore have an event horizon that is not also a Killing horizon\footnote{This does not violate the rigidity theorems since our solutions do not have compact horizons (see e.g. \cite{Hollands:2006rj}).}.

\end{section}

\vskip 1cm
\centerline{\bf Acknowledgements}
\vskip .5 cm
It is a pleasure to thank Veronika Hubeny, Mukund Rangamani,Toby Wiseman, and most especially Donald Marolf for helpful discussions. We thank the organizers and participants of the program ``New perspectives from strings and higher dimensions", Centro de Ciencias de Benasque Pedro Pascual (2011) for stimulating discussions. This work was supported in part by NSF Grants PHY08-55415 and PHY12-05500.

\appendix
\begin{section}{\label{appendix:a}Asymptotic expansion}
In this appendix we present the functions used in section 3.  The functions for the Schwarzschild funnel appearing in \eqref{eqs:asymptoticexpansion}, \eqref{eq:tofeffer} and \eqref{eq:holostressschwarzschild} are:
\begin{subequations}
\begin{multline}
L_T(x) = 4185 x^{11}-8409 x^{10}-13770 x^9+44406 x^8-16301 x^7-40548 x^6+37538 x^5\\+2401 x^4-15024 x^3+6411 x^2-913 x+26\,,
\end{multline}
\begin{equation}
L_A(x) = -81 x^9+418 x^8-1037 x^7+1336 x^6-466 x^5-840 x^4+1002 x^3-356 x^2+23 x+2\,,
\end{equation}
\begin{equation}
L_F(x) = -27 x^4+31 x^3+22 x^2-30 x+5\,,
\end{equation}
\begin{multline}
L_S(x) = 5400 x^{11}-11460 x^{10}-14496 x^9+51308 x^8-18576 x^7-46818 x^6+41490 x^5\\+4333 x^4-17163 x^3+6892 x^2-939 x+32\,,
\end{multline}
\begin{equation}
R_4(\tilde{x})=-\frac{9 \tilde{x}^6-30 \tilde{x}^5+31 \tilde{x}^4-4 \tilde{x}^3-9 \tilde{x}^2+\tilde{x}+2}{2 \tilde{x}^2 \left(\tilde{x}+1\right)^4}\,,
\end{equation}
\begin{multline}
R_6(\tilde{x}) = [8 b_2(\tilde{x})-\tilde{x}^2 (2349 \tilde{x}^{10}-10644 \tilde{x}^9+15150 \tilde{x}^8+644 \tilde{x}^7-21817 \tilde{x}^6+18146 \tilde{x}^5\\
+1342\tilde{x}^4-9176 \tilde{x}^3+5023 \tilde{x}^2-1114 \tilde{x}+95)] /[32 \tilde{x}^3 (1-\tilde{x}^2)^4]\,,
\end{multline}
\begin{equation}
V_4(\tilde{x}) = \frac{(1-\tilde{x})^4 (72 \tilde{x}^6-64 \tilde{x}^5-176 \tilde{x}^4+192 \tilde{x}^3+40 \tilde{x}^2-63 \tilde{x}-3)}{4 (1+\tilde{x})^3}\,,
\end{equation}
\begin{multline}
\tilde{T}_{tt} = -16443 \tilde{x}^{12}+48522 \tilde{x}^{11}+25167 \tilde{x}^{10}-233180 \tilde{x}^9+234445 \tilde{x}^8+107732 \tilde{x}^7\\
-313203 \tilde{x}^6+132690 \tilde{x}^5+75637 \tilde{x}^4-86532\tilde{x}^3+28578 \tilde{x}^2-3532 \tilde{x}+95\,,
\end{multline}
\begin{multline}
\tilde{T}_{\tilde{x}\tilde{x}} =2349 \tilde{x}^{10}-10644 \tilde{x}^9+15150 \tilde{x}^8+644 \tilde{x}^7-21817 \tilde{x}^6+18146 \tilde{x}^5\\
+1342 \tilde{x}^4-9176 \tilde{x}^3+5023 \tilde{x}^2-1114 \tilde{x}+95\,,
\end{multline}
\begin{multline}
\tilde{T}_{\Omega} =-11745 \tilde{x}^{12}+37878 \tilde{x}^{11}+1677 \tilde{x}^{10}-138028 \tilde{x}^9+156879 \tilde{x}^8+48786 \tilde{x}^7\\
-187669 \tilde{x}^6+87840 \tilde{x}^5+40131 \tilde{x}^4-51206\tilde{x}^3+17772 \tilde{x}^2-2418 \tilde{x}+95\,.
\end{multline}
\end{subequations}

Now for the four-dimensional funnel, we give the expressions for the functions appearing in \eqref{eqs:asymptoticexpansion2}, \eqref{eq:tofeffer2}.  They are given by
\begin{subequations}
\be
K_T(x)=4 \big[1 + x (2 + 9 x^3 (1 - x^2)^4)\big] + (1 - x^2) (1 + x)^2 H_-(x)
\ee
\begin{multline}
K_F(x)=24(1-x)^2(1+2x)+3(1-x^2)^3H_+(x)-6(1-x)x^2(1+x)^2g(x)^3\\
-\frac{(1-x)x^2(1+x)^2(1-x^2)\lambda^2J(x)g(x)^2}{\ell(x)^2}+\left[-\frac{3x^2(1-x)^2p_1(x)}{\ell(x)}+3p_2(x)\right]g(x)
\end{multline}
\be
K_B(x)=4p_3(x)-(1-x^2)(1+x)^2H_-(x)
\ee
\be
K_S(x)=12(1-x^2)^2x^2\big[6(1-x^2)^2x^2+(1+x)J(x)\big]-K_T(x)
\ee
\be
Q_3(x)=-\frac{4(1-x)^2\big[1+x(2+9x^3(1-x^2)^4)\big]+(1-x^2)^4g(x)\big[12(1-x)x^2+g(x)\big]}{16x^3g(x)^3}
\ee
\be
U_4(x)=-\frac{(1-x)(1-x^2)^4}{128x^3g(x)^3}\Big\{4p_4(x)+(1-x)\big[6(1-x)x^2+g(x)\big]K_B(x)+2x^2K_F(x)\Big\}\;,
\ee
\end{subequations}
where

\begin{subequations}
\be
H_\pm(x)=\frac{g(x)^2}{\ell(x)^2}\Big[x^4(1-x^2)\lambda^2\pm(1+3x^2)\ell(x)^2\Big]
\ee
\be
J(x)=\frac{g(x)}{\ell(x)}\Big[x^2(1-x^2)\lambda+(1-3x^2)\ell(x)\Big]
\ee
\be
p_1(x)=32x^5-43x^4-17x^3-8x^2-2x+8
\ee
\be
p_2(x)=18 x^{11} - 57 x^{10} + x^9 - 38 x^8 - 112 x^7 + 327 x^6 + 157 x^5 - 
 316 x^4 - 72 x^3 + 88 x^2 + 8 x - 2
\ee
\begin{multline}
p_3(x)=-33 x^{12} + 15 x^{11} + 126 x^10 + 15 x^9 - 114 x^8 - 81 x^7 - 30 x^6 + 
 57 x^5 + 63 x^4\\ - 6 x^3 - 12 x^2 - 2 x - 1
 \end{multline}
 \be
 p_4(x)=6 (1 - x)^2 x^2 (1 + 2 x) + 
 \Big\{1 + 
    x (1 + x) \Big[1 + (1 - x)^2 x (1 + x)^2 (2 + x^2 (9 - 8 x))^2\Big]\Big\} g(x)\;,
 \ee
 and $g$ and $\ell$ were already defined in \eqref{fgell}.  
\end{subequations}
\end{section}

\singlespacing

\end{document}